\documentstyle[preprint,aps]{revtex}

\tightenlines

\catcode`@=11
\def\references{%
\ifpreprintsty
\bigskip\bigskip
\hbox to\hsize{\hss\large \refname\hss}%
\else
\vskip24pt
\hrule width\hsize\relax
\vskip 1.6cm
\fi
\list{\@biblabel{\arabic{enumiv}}}%
{\labelwidth\WidestRefLabelThusFar  \labelsep4pt %
\leftmargin\labelwidth %
\advance\leftmargin\labelsep %
\ifdim\baselinestretch pt>1 pt %
\parsep  4pt\relax %
\else %
\parsep  0pt\relax %
\fi
\itemsep\parsep %
\usecounter{enumiv}%
\let\p@enumiv\@empty
\def\theenumiv{\arabic{enumiv}}%
}%
\let\newblock\relax %
\sloppy\clubpenalty4000\widowpenalty4000
\sfcode`\.=1000\relax
\ifpreprintsty\else\small\fi
}
\catcode`@=12

\newcommand{\be}{\begin{equation}}
\newcommand{\ee}{\end{equation}}

\newcommand{\eq}[1]{eq.~(\ref{#1})}

\newcommand{\chii}{\chi_{{}_{\rm I}}}
\newcommand{\chir}{\chi_{{}_{\rm R}}}
\newcommand{\pbar}{\bar p}

\begin{document}
\preprint{
\vbox{
\hbox{\bf NUHEP 712}
\hbox{November 30, 2000}}}
\title{\vspace*{.5in}
Consequences of the Factorization Hypothesis in $\pbar p,\  pp,\ \gamma p\ {\rm and \ }\gamma\gamma$ Collisions
}
\author{
M.~M.~Block}
\address{ Department of Physics and Astronomy,
Northwestern University, Evanston, IL 60208, USA}
\author
{A.~B.~Kaidalov}
\address{ITEP,
B. Cheremushkinskaya ulitsa 25
117259 Moscow, Russia}
\maketitle
\begin{abstract}
Using an eikonal analysis, we examine the validity of the factorization theorem  for nucleon-nucleon, $\gamma p$ and $\gamma\gamma$ collisions. As an example, using the additive quark model and meson vector dominance, we directly show that for all energies and values of the eikonal, that the factorization theorem $\sigma_{\rm nn}/\sigma_{\gamma p}=\sigma_{\gamma p}/\sigma_{\gamma \gamma}$ holds. We can also compute the survival probability of large rapidity gaps in high energy $\pbar p$ and $pp$ collisions. We show that the survival probabilities are identical (at the {\em same} energy) for $\gamma p$ and $\gamma \gamma$ collisions, as well as for nucleon-nucleon collisions. We further show that neither the factorization theorem nor the reaction-independence of the survival probabilities depends on the assumption of an additive quark model, but, more generally, depends on the {\em opacity} of the eikonal being {\em independent} of whether the reaction is n-n, $\gamma p$ or $\gamma \gamma$. 
\end{abstract}
\thispagestyle{empty}
\newpage
%
%
%%%%%% Start text
In this note, we will use an eikonal model to make  calculations of cross sections and survival probabilities of rapidity gaps in nucleon-nucleon, $\gamma p$ and $\gamma \gamma$ collisions. 

In an eikonal model\cite{blockhalzenpancheri}, we define our (complex) eikonal $\chi(b,s)$ so that $a(b,s)$, the (complex) scattering amplitude in impact parameter space $b$, is given by
\be
a(b,s)=\frac{i}{2}\left(1-e^{i\chi(b,s)}\right)=\frac{i}{2}\left(1-e^{-\chii(b,s)+i\chir(b,s)}\right).\label{eik}
\ee
Using the optical theorem,
the total cross section $\sigma_{\rm tot}(s)$ is given by 
\begin{equation}
\sigma_{\rm tot}(s)=2\int\,\left[1-e^{-\chii (b,s)}\cos(\chir(b,s))\right]\,d^2\vec{b},\label{sigtot}
\end{equation}
the elastic scattering cross section 
$\sigma_{\rm el}(s)$ is given by
\begin{eqnarray}
\sigma_{\rm elastic}(s)&=&\int\left|1-e^{-\chii(b,s)+i\chir(b,s)}\right|^2\,d^2\vec{b}\label{sigel}
\end{eqnarray}
and
the inelastic cross section, $\sigma_{\rm inelastic}(s)$, is given by
\be
\sigma_{\rm inelastic}(s)=\sigma_{\rm tot}(s)-\sigma_{\rm elastic}(s)=\int\,\left [1-e^{-2\chii(b,s)}\right ]\,d^2\vec{b}.\label{sigin}
\ee
The ratio of the real to the imaginary portion of the forward nuclear scattering amplitude, $\rho$,
is given by  
\begin{eqnarray}
\rho(s)&=&\frac{{\rm Re}\left\{i(\int 1-e^{-\chii(b,s)+i\chir(b,s)})\,d^2\vec{b}\right\}}
{{\rm Im}\left\{i(\int (1-e^{-\chii(b,s)+i\chir(b,s)})\,d^2\vec{b}\right\}}\label{rho}
\end{eqnarray}
and the nuclear slope parameter $B$ is given by
\be
B=
\frac{\int\,b^2a(b,s)\,d^2\vec{b}}{2\int\,a(b,s)\,d^2\vec{b}}.\label{Bsimple}
\ee

It is easily shown\cite{bjorken}, using  \eq{sigin}, that the differential probability in impact parameter space $b$, for {\em not} having an inelastic interaction, is given by 
\be
P_{\rm no\  inelastic}=e^{-2\chii(b,s)}.\label{noinelastic}
\ee

Block {\em et al.}\cite{blockhalzenpancheri} have used an even QCD-Inspired eikonal $\chi_{\rm even}$ given by the sum of three contributions, glue-glue, quark-glue and quark-quark, which are individually factorizable into a product of a cross section  $\sigma (s)$ times an impact parameter space distribution function $W(b\,;\mu)$,  {\em i.e.,}:
\begin{eqnarray}
 \chi^{\rm even}(s,b)& = &\chi_{\rm gg}(s,b)+\chi_{\rm qg}(s,b)+\chi_{\rm qq}(s,b)\nonumber\\
&=&i\left[\sigma_{\rm gg}(s)W(b\,;\mu_{\rm gg})+\sigma_{\rm qg}(s)W(b\,;\mu_{\rm qg})+\sigma_{\rm qq}(s)W(b\,;\mu_{\rm qq})\right],\label{eq:chieven}
\end{eqnarray}
where we have set $\mu_{\rm qg}=\sqrt{\mu_{\rm qq}\mu_{\rm gg}}$ and
where the impact parameter space distribution function 
\begin{equation}
W(b\,;\mu)=\frac{\mu^2}{96\pi}(\mu b)^3K_3(\mu b)\label{W}
\end{equation}
is normalized so that
%\begin{equation}
$\int W(b\,;\mu)d^2 \vec{b}=1.$  
%\end{equation}
Hence, the $\sigma$'s in \eq{eq:chieven} have the dimensions of a cross section. The factor $i$ is inserted in \eq{eq:chieven} since the high energy eikonal is largely imaginary (the $\rho$ value for nucleon-nucleon scattering is rather small).   
The total even contribution is not yet analytic. 
For large $s$, the {\rm even} amplitude in \eq{eq:chieven} is made analytic by the substitution $s\rightarrow se^{-i\pi/2}$ (see  the table on p. 580 of reference  \cite{bc}, along with reference \cite{eden}).  The quark contribution $\chi_{\rm qq}(s,b)$ accounts for the constant cross section and a Regge descending component ($\propto 1/\sqrt s$), whereas the mixed quark-gluon term $\chi_{\rm qg}(s,b)$ simulates diffraction ($\propto \log s$).  The glue-glue term $\chi_{\rm gg}(s,b)$, which eventually rises as a power law  $s^\epsilon$,  accounts for the rising cross section and dominates at the highest energies. In \eq{eq:chieven}, the inverse sizes (in impact parameter space) $\mu_{\rm qq}$ and $\mu_{\rm gg}$ are to be fit by experiment, whereas the quark-gluon inverse size is taken as $\mu_{\rm qg}=\sqrt{\mu_{\rm qq}\mu_{\rm gg}}$.   

The high energy analytic {\em odd} amplitude (for its structure in $s$, see eq. (5.5b) of reference \cite{bc}, with $\alpha =0.5$)
that fits the data is given by
\begin{eqnarray}
\chii^{\rm odd}(b,s)&=&-\sigma_{\rm odd}\,W(b;\mu_{\rm odd}),\label{oddanalytic}
\end{eqnarray}
with $\sigma_{\rm odd}\propto 1/\sqrt s$, and 
with 
%\be
$W(b,\mu_{\rm odd})=\frac{\mu_{\rm odd}^2}{96\pi}(\mu_{\rm odd} b)^3\,K_3(\mu_{\rm odd} b)$ %\label{Woddnormalization},
%\ee
normalized so that
%\begin{equation}
$\int W(b\,;\mu_{\rm odd})d^2 \vec{b}=1.$% \label{oddWintegral}
%\end{equation}
%Hence, the $\sigma_{\rm odd}$ in \eq{oddanalytic} has the dimensions of a cross section.

Finally, 
\be
\chi^{\pbar p}_{pp}=\chi_{\rm even}\pm \chi_{\rm odd}.\label{totalchi}
\ee

The eikonal of \eq{eq:chieven}
is a QCD-inspired parameterization of the forward
 proton--proton and proton--antiproton scattering amplitudes\,\cite{blockhalzenpancheri}
 which is analytic, unitary, satisfies crossing symmetry, the Froissart Bound (asymptotically approaching a black disk) and, using a $\chi^2$ procedure, fits all accelerator data of  
$\sigma_{\rm tot}$ (including the new E-811 Tevatron cross section\cite{orear}), nuclear slope parameter $B$
 and $\rho$, the ratio of the real-to-imaginary part of the forward
 scattering amplitude for both $pp$ and $\pbar p$ collisions. In addition, the high energy cosmic ray cross sections of Fly's Eye\,\cite{fly} and  
AGASSA\,\cite{akeno} experiments are also simultaneously used\cite{blockcr},  dramatically reducing the errors of the fitted parameters. The vast wealth of data fitting a myriad of $\gamma p$ and $\gamma \gamma$ collisions\cite{blockhalzenpancheri} as well as nucleon-nucleon data furnishes an empirical justification for the model---% 
for lack of space, these results are not reproduced here (see ref. \cite{blockhalzenpancheri} and ref. \cite{blockcr} for more detail), since the main thrust of this note is to show the consequences of the factorization hypothesis on rapidity gaps and on the ratios of cross sections for the processes nn, $\gamma p$ and $\gamma\gamma$.

As an example of a large rapidity gap process, we consider the production cross section for Higgs-boson production through W fusion.  The inclusive differential cross section in impact parameter space $b$ is given by
$\frac{d\sigma}{d^2\vec{b}}=\sigma_{{\rm WW}\rightarrow {\rm H}}\,W(b\,;\mu_{\rm qq}),$ % \label{eq:xsection}
where we have assumed that $W(b\,;\mu_{\rm qq})$ (the differential impact parameter space {\em quark} distribution in the proton) is the same as that of the W bosons.  

The cross section for producing the Higgs boson {\em and} having a large rapidity gap (no secondary particles) is given by
\be
\frac{d\sigma_{\rm gap}}{d^2\vec{b}}=\sigma_{{\rm WW}\rightarrow {\rm H}}\,W(b\,;\mu_{\rm qq})e^{-2\chii(s,b)}=\sigma_{{\rm WW}\rightarrow {\rm H}}\,\frac{d(|S|^2)}{d\vec{b}^2}. \label{eq:nosecondaries}
\ee

A generalization of this result is to define $<|S|^2>$, the survival probability of {\em any} large rapidity gap\cite{bjorken,maor}, as
\be 
<|S|^2>=\int W(b\,;\mu_{\rm qq})e^{-2\chii(s,b)}d^2\,\vec{b},\label{eq:survival}
\ee 
where for $\chii(s,b)$, we will use $\chii(s,b)^{\rm even}$ given by \eq{eq:chieven}, {\em i.e.,} we will neglect the small differences between $\pbar p$ and $pp$ collisions at low energy, in effect averaging them.  
We note that the energy dependence of the survival probability $<|S|^2>$ is through the energy dependence of $\chii$, the imaginary portion of the eikonal. 

In ref. \cite{blockhalzenpancheri}, using the additive quark model and vector dominance, the authors assume that the eikonal $\chi^{\gamma p}$ for $\gamma p$ reactions is determined by substituting $\sigma\rightarrow \frac{2}{3}\sigma$, $\mu\rightarrow\sqrt{\frac{3}{2}}\mu$ into $\chi^{\rm even}(s,b)$, given by \eq{eq:chieven}. In turn, they determine $\chi^{\gamma \gamma}$ for $\gamma\gamma$ reactions by substituting $\sigma\rightarrow \frac{2}{3}\sigma$, $\mu\rightarrow\sqrt{\frac{3}{2}}\mu$ into $\chi^{\gamma p}(s,b)$. 
The naive additive quark model and vector dominance suggest that $\sigma\rightarrow \frac{2}{3}\sigma$, since there are 2 quarks in a photon and 3 in a nucleon.  Their transformation $\mu\rightarrow\sqrt{\frac{3}{2}}\mu$ is justified as follows.  Let us require that the ratio of elastic to total scattering be process-independent, {\em i.e.,}
\be
\left(\frac{\sigma_{\rm elastic}}{\sigma_{\rm tot}}\right)^{\rm nn}=
\left(\frac{\sigma_{\rm elastic}}{\sigma_{\rm tot}}\right)^{\gamma p}=
\left(\frac{\sigma_{\rm elastic}}{\sigma_{\rm tot}}\right)^{\rm \gamma\gamma}\label{eq:ratio}
\ee
at {\em all} energies, a condition that insures that each process becomes equally black disk-like as we go to high energy.  
For simplicity, we will evaluate $\left(\frac{\sigma_{\rm elastic}}{\sigma_{\rm tot}}\right)^{\rm nn}$ in the small eikonal limit, utilizing \eq{sigel}, \eq{sigtot} and \eq{W}, using for our eikonal the toy version $\chi^{\rm nn}(s,b) =i(\sigma_{gg}W(b;\mu_{gg}))$.  We find that 
\be
\left(\frac{\sigma_{\rm elastic}}{\sigma_{\rm tot}}\right)^{\rm nn}= \sigma_{gg}\mu_{gg}^2\times(\frac{1}{96\pi})^2\int y^6(K_3(y))^2d^2\vec y,\quad {\rm where\ } y=\mu_{gg} b.
\ee
Thus, for the ratio to be process-independent, 
\begin{equation}
(\mu_{gg})^{\gamma \gamma}=\sqrt{\frac{3}{2}}(\mu_{gg})^{\gamma p}=\frac{3}{2}(\mu_{gg})^{\rm nn},\quad 
{\rm since}\label{eq:mus}\end{equation}
\be
(\sigma_{gg})^{\gamma \gamma}=\frac{2}{3}(\sigma_{gg})^{\gamma p}=\frac{4}{9}(\sigma_{gg})^{\rm nn}\label{eq:sigmas}.
\ee
This argument is readily generalized to {\em all} $\mu$, leading to {\em each} $\sigma \mu^2$ being {\em process-independent}. 

Indeed, the consequences of \eq{eq:ratio} that each $\sigma \mu^2$ is  process-independent can be restated more simply in the following language:
\begin{itemize}
\item We  require that the eikonal of \eq{eq:chieven} have the {\em same opacity} for n-n, $\gamma p$ and $\gamma \gamma$ scattering,\end{itemize} 
where the opacity is the value of the eikonal at $b=0$. 

For specificity, however, we will use the eikonal of \eq{eq:chieven}, with the conditions of \eq{eq:mus} and \eq{eq:sigmas}, hereafter.

Thus, 
\be
\chi^{\gamma p}(s,b)= i\left[\frac{2}{3}\sigma_{\rm gg}(s)W(b\,;\sqrt{\frac{3}{2}}\mu_{\rm gg})+\frac{2}{3}\sigma_{\rm qg}(s)W(b\,;\sqrt{\frac{3}{2}}\mu_{\rm qg})+\frac{2}{3}\sigma_{\rm qq}(s)W(b\,;\sqrt{\frac{3}{2}}\mu_{\rm qq})\right], \label{eq:chigp}
\ee
and
\be
\chi^{\gamma \gamma}(s,b)= i\left[\frac{4}{9}\sigma_{\rm gg}(s)W(b\,;\frac{3}{2}\mu_{\rm gg})+\frac{4}{9}\sigma_{\rm qg}(s)W(b\,;\frac{3}{2}\mu_{\rm qg})+\frac{4}{9}\sigma_{\rm qq}(s)W(b\,;\frac{3}{2}\mu_{\rm qq})\right]. \label{eq:chigg}
\ee
Since the normalization of each $W(b;\mu)$ above is proportional to $\mu^2$, it is easy to see, using the new dimensionless variable $x_q=\sqrt{\frac{3}{2}}\mu_{\rm qq}b$
 that
\begin{eqnarray}
\chi^{\gamma p}(s,b)&=&\frac{i}{96\pi}\left[\sigma_{\rm gg}\mu_{\rm gg}^2\left(\frac{\mu_{\rm gg}}{\mu_{\rm qq}}x_q\right)^3K_3(\frac{\mu_{\rm gg}}{\mu_{\rm qq}}x_q)+\sigma_{\rm qg}\mu_{\rm qg}^2\left(\frac{\mu_{\rm qg}}{\mu_{\rm qq}}x_q\right)^3K_3(\frac{\mu_{\rm qg}}{\mu_{\rm qq}}x_q)\right.\nonumber\\
&&\ \ \quad \quad \quad +\left. \vphantom{\frac{\sqrt{ \mu_{\rm gg}\mu_{\rm qq}}}{\mu_{\rm qq}}}\sigma_{\rm qq}\mu_{\rm qq}^2(x_q)^3K_3(x_q)\right].\label{eq:newchigp}
\end{eqnarray}
Thus,
\begin{eqnarray}
<|S^{\gamma p}|^2>&=&\frac{1}{96\pi}\int x_q^3K_3(x_q)\times \nonumber\\
&&{\rm exp}-\frac{1}{48\pi}\left[\sigma_{\rm gg}\mu_{\rm gg}^2\left(\frac{\mu_{\rm gg}}{\mu_{\rm qq}}x_q\right)^3K_3(\frac{\mu_{\rm gg}}{\mu_{\rm qq}}x_q)+\sigma_{\rm qg} \mu_{\rm qg}^2\left(\frac{\mu_{\rm qg}}{\mu_{\rm qq}}x_q\right)^3K_3(\frac{\mu_{\rm qg}}{\mu_{\rm qq}}x_q)\right.\nonumber\\
&&\ \ \quad \quad \quad +\left. \vphantom{\frac{\sqrt{ \mu_{\rm gg}\mu_{\rm qq}}}{\mu_{\rm qq}}}\sigma_{\rm qq}\mu_{\rm qq}^2(x_q)^3K_3(x_q)\right]\,d^2\vec{x_q}
\label{eq:Sgp}
\end{eqnarray} 
and 
\begin{eqnarray}
<|S^{\gamma \gamma}|^2>&=&\frac{1}{96\pi}\int x_g^3K_3(x_g)\times \nonumber\\
&&{\rm exp}-\frac{1}{48\pi}\left[\sigma_{\rm gg}\mu_{\rm gg}^2\left(\frac{\mu_{\rm gg}}{\mu_{\rm qq}}x_g\right)^3K_3(\frac{\mu_{\rm gg}}{\mu_{\rm qq}}x_g)+\sigma_{\rm qg} \mu_{\rm qg}^2\left(\frac{\mu_{\rm qg}}{\mu_{\rm qq}}x_g\right)^3K_3(\frac{\mu_{\rm qg}}{\mu_{\rm qq}}x_g)\right.\nonumber\\
&&\ \ \quad \quad \quad +\left. \vphantom{\frac{\sqrt{ \mu_{\rm gg}\mu_{\rm qq}}}{\mu_{\rm qq}}}\sigma_{\rm qq}\mu_{\rm qq}^2(x_g)^3K_3(x_g)\right]\,d^2\vec{x_g}
\label{eq:Sgg}
\end{eqnarray} 
where we used the variable substitution $x_g=\frac{3}{2}\mu_{\rm qq}b$.
Finally, we have, using the variable substitution $x_n=\mu_{\rm qq}b$,
\begin{eqnarray}
<|S^{\rm even}|^2>&=&\frac{1}{96\pi}\int x_n^3K_3(x_n)\times \nonumber\\
&&{\rm exp}-\frac{1}{48\pi}\left[\sigma_{\rm gg}\mu_{\rm gg}^2\left(\frac{\mu_{\rm gg}}{\mu_{\rm qq}}x_n\right)^3K_3(\frac{\mu_{\rm gg}}{\mu_{\rm qq}}x_n)+\sigma_{\rm qg}\mu_{\rm qg}^2\left(\frac{\mu_{\rm qg}}{\mu_{\rm qq}}x_n\right)^3K_3(\frac{\mu_{\rm qg}}{\mu_{\rm qq}}x_n)\right.\nonumber\\
&&\ \ \quad \quad \quad +\left. \vphantom{\frac{\sqrt{ \mu_{\rm gg}\mu_{\rm qq}}}{\mu_{\rm qq}}}\sigma_{\rm qq}\mu_{\rm qq}^2(x_n)^3K_3(x_n)\right]\,d^2\vec{x_n}.
\label{eq:Seven}
\end{eqnarray} 
Thus, comparing \eq{eq:Sgp}, \eq{eq:Sgg} and \eq{eq:Seven}, we see that 
\be 
<|S^{\gamma p}|^2>=<|S^{\gamma \gamma}|^2>=<|S^{\rm even}|^2>.\label{eq:allequal}
\ee
We see from \eq{eq:allequal} that $<|S|^2>$, the survival probability for nucleon-nucleon, $\gamma p$ and $\gamma \gamma$ collisions, is {\em reaction-independent}, depending {\em only} on $\sqrt{s}$, the cms energy of the collision.  We emphasize that this result is much more general, being true for {\em any} eikonal whose opacity is process-independent---not only for the additive quark model that we have employed.  

In order to calculate the total nucleon-nucleon cross section, we use the variable substitution $x_n=\mu_{\rm qq}b$ and rewrite $\chi^{\rm even}$ of \eq{eq:chieven} as  
\begin{eqnarray}
\chi^{\rm even}(s,b)&=&\frac{i}{96\pi}\left[\sigma_{\rm gg}\mu_{\rm gg}^2(\frac{\mu_{\rm gg}}{\mu_{\rm qq}}x_n)^3K_3(\frac{\mu_{\rm gg}}{\mu_{\rm qq}}x_n)+\sigma_{\rm qg}\mu_{\rm qg}^2\left(\frac{\mu_{\rm qg}}{\mu_{\rm qq}}x_n\right)^3K_3(\frac{\mu_{\rm qg}}{\mu_{\rm qq}}x_n)\right.\nonumber\\
&&\ \ \quad \quad \quad +\left. \vphantom{\frac{\sqrt{ \mu_{\rm gg}\mu_{\rm qq}}}{\mu_{\rm qq}}}\sigma_{\rm qq}\mu_{\rm qq}^2(x_n)^3K_3(x_n)\right].\label{eq:newchinn}
\end{eqnarray}
Using \eq{sigtot} and approximating $\chi_{\rm even}$ in \eq{eq:newchinn} as pure imaginary, we have 
\begin{eqnarray}
\sigma_{\rm tot}^{\rm nn}(s)&=&2\int \left(1-{\rm exp}-\frac{1}{96\pi}\left[\sigma_{\rm gg}\mu_{\rm gg}^2(\frac{\mu_{\rm gg}}{\mu_{\rm qq}}x_n)^3K_3(\frac{\mu_{\rm gg}}{\mu_{\rm qq}}x_n)+\sigma_{\rm qg}\mu_{\rm qg}^2\left(\frac{\mu_{\rm qg}}{\mu_{\rm qq}}x_n\right)^3K_3(\frac{\mu_{\rm qg}}{\mu_{\rm qq}}x_n)\right.\right.\nonumber\\
&&\ \ \quad \quad \quad +\left. \left.\vphantom{\frac{\sqrt{ \mu_{\rm gg}\mu_{\rm qq}}}{\mu_{\rm qq}}}\sigma_{\rm qq}\mu_{\rm qq}^2(x_n)^3K_3(x_n)\right]\right)\frac{1}{\mu_{\rm qq}^2}\,d^2\vec{x_n}.\label{eq:signn}
\end{eqnarray}
Using Vector Meson Dominance and the additive quark model and letting $P_{\rm had}^\gamma$ be the probability that a $\gamma$ ray materialize as a hadron, we find, using $\chi^{\gamma p}$ from \eq{eq:newchigp}, that
\begin{eqnarray}
\sigma_{\rm tot}^{\gamma p}(s)&=&2\int \left(1-{\rm exp}-\frac{1}{96\pi}\left[\sigma_{\rm gg}\mu_{\rm gg}^2(\frac{\mu_{\rm gg}}{\mu_{\rm qq}}x_q)^3K_3(\frac{\mu_{\rm gg}}{\mu_{\rm qq}}x_q)+\sigma_{\rm qg}\mu_{\rm qg}^2\left(\frac{\mu_{\rm qg}}{\mu_{\rm qq}}x_q\right)^3K_3(\frac{\mu_{\rm qg}}{\mu_{\rm qq}}x_q)\right.\right.\nonumber\\
&&\ \ \quad \quad \quad +\left. \left.\vphantom{\frac{\sqrt{ \mu_{\rm gg}\mu_{\rm qq}}}{\mu_{\rm qq}}}\sigma_{\rm qq}\mu_{\rm qq}^2(x_q)^3K_3(x_q)\right]\right)\frac{2}{3P_{\rm had}^\gamma }\frac{1}{\mu_{\rm qq}^2}\,d^2\vec{x_q},\label{eq:siggp}
\end{eqnarray}
where $x_q=\sqrt{\frac{3}{2}}\mu_{\rm qq}b$.  Finally, substituting $x_g=\frac{3}{2}\mu_{\rm qq}b$ into \eq{eq:chigp} and using \eq{sigtot}, we evaluate $\sigma_{\rm tot}^{\gamma \gamma}$ as
\begin{eqnarray}
\sigma_{\rm tot}^{\gamma \gamma}(s)&=&2\int \left(1-{\rm exp}-\frac{1}{96\pi}\left[\sigma_{\rm gg}\mu_{\rm gg}^2(\frac{\mu_{\rm gg}}{\mu_{\rm qq}}x_g)^3K_3(\frac{\mu_{\rm gg}}{\mu_{\rm qq}}x_g)+\sigma_{\rm qg}\mu_{\rm qg}^2\left(\frac{\mu_{\rm qg}}{\mu_{\rm qq}}x_g\right)^3K_3(\frac{\mu_{\rm qg}}{\mu_{\rm qq}}x_g)\right.\right.\nonumber\\
&&\ \ \quad \quad \quad +\left. \left.\vphantom{\frac{\sqrt{ \mu_{\rm gg}\mu_{\rm qq}}}{\mu_{\rm qq}}}\sigma_{\rm qq}\mu_{\rm qq}^2(x_g)^3K_3(x_g)\right]\right){\left(\frac{2}{3P_{\rm had}^\gamma }\right)}^2\frac{1}{\mu_{\rm qq}^2}\,d^2\vec{x_g}.\label{eq:siggg}
\end{eqnarray}
 Clearly, from inspection of \eq{eq:signn}, \eq{eq:siggp} and \eq{eq:siggg}, we see that the factorization theorem
\be
\frac{\sigma_{\rm tot}^{\rm nn}(s)}{\sigma_{\rm tot}^{\gamma p}(s)}=\frac{\sigma_{\rm tot}^{\gamma p}(s)}{\sigma_{\rm tot}^{\gamma \gamma}(s)}\label{eq:factorization}
\ee
holds at all energies, {\em i.e.}, the factorization theorem survives exponentiation. We again emphasize  that 
this result is true for {\em any} eikonal which factorizes into sums of $\sigma_i (s)\times W_i(b;\mu)$ having the scaling feature that the product $\sigma_i\mu_i^2$ is reaction-independent---not only for the additive quark model that we have employed here, but for {\em any} eikonal whose opacity is independent of whether the reaction is n-n, $\gamma p$ or $\gamma\gamma$.  It is valid at {\em all} energies, independent of the size of the eikonal and independent of the details of the initial factorization scheme. In the particular scheme of \eq{eq:signn}, \eq{eq:siggp} and \eq{eq:siggg}, {\em i.e.,} for Vector Meson Dominance and the Additive Quark Model, the proportionality constant is $2/(3P_{\rm had}^\gamma)$.  Clearly, the same proportionality constant holds for both $\sigma_{\rm elastic}$ and $\sigma_{\rm inelastic}$, whereas for the nuclear slope parameter $B$ it is easily shown that the proportionality constant is 2/3.  For $\rho$, the proportionality constant is unity. 

We note that in the Regge approach the functions $W_i(b,\mu)$ depend on $s$,
since the radius of interaction ($R\sim 1
/\mu$) for the Pomeron exchange which determines
an eikonal grows with energy as $R^2=R_0^2+\alpha_P^{\prime} ln(s/s_0)$. Thus, in the Regge
eikonal model, factorization breaks down in general. However, the Pomeron slope 
$\alpha_P^{\prime}$ is very small ($\approx~0.2~GeV^{-2}$) and the $R_{0}$ for 
different processes are approximately in the same relation as in the model discussed
above. As a result, the factorization relations are valid with  good accuracy even in a Regge eikonal
model, in a broad region of energies\cite{kane,review}.  In the model discussed earlier, where the opacity was independent of the reaction, the values of $\mu$ were chosen to be energy independent and the factorization relations of \eq{eq:factorization} are exact at all energies.

The differential scattering cross section for elastic nucleon-nucleon scattering can be written as
\be
\frac{d\sigma^{\rm nn}}{dt}(s,t)=\frac{1}{4\pi}\left|\int J_0(qb)
\left(1-e^{i\chi^{\rm even}(b,s)}\right)\,d^2\vec{b}\,\right|^2\; ,
\label{eq:dsdt2}
\ee
where the squared 4-momentum transfer $t=-q^2$. 
It is straightforward, by appropriate variable transformation, to show that the differential cross section for the ``elastic" scattering reaction $\gamma +p\rightarrow V+p$ is given by
\be
\frac{d\sigma_{Vp}^{\gamma p}}{dt}(s,t)=\frac{4}{9P_V^\gamma}\frac{d\sigma^{\rm nn}}{dt}(s,\frac{2}{3}t),\label{eq:dsdtgp}
\ee
where $V$ is the vector meson $\rho$, $\omega$ or $\phi$ and $P_V^\gamma$ is the probability that a photon goes into the vector meson $V$.
For the ``elastic" scattering reaction $\gamma +\gamma\rightarrow V+V$,  we can show that
\be
\frac{d\sigma_{VV}^{\gamma \gamma}}{dt}(s,t)={\left(\frac{4}{9P_V^\gamma}\right)}^2\frac{d\sigma^{\rm nn}}{dt}(s,\frac{4}{9}t).\label{eq:dsdtgg}
\ee
Thus, a knowledge of $\frac{d\sigma^{\rm nn}}{dt}(s,t)$ for elastic nucleon-nucleon scattering determines the differential ``elastic'' scattering cross sections for the reactions $\gamma +p\rightarrow V+p$ and $\gamma +\gamma\rightarrow V+V$.
We now write the factorization theorem for differential elastic scattering as
\be\frac{d\sigma^{\rm nn}}{dt}(s,t){ \  \Bigg / \ }\frac{d\sigma_{Vp}^{\gamma p}}{dt}(s,{\frac{3}{2}}t)
=\frac{d\sigma_{Vp}^{\gamma p}}{dt}(s,{\frac{3}{2}}t){ \ \Bigg /\ }\frac{d\sigma_{VV}^{\gamma \gamma}}{dt}(s,{\frac{9}{4}}t).\label{eq:dsdtfactorize}
\ee

The survival probabilities $<|S|^2>$ are independent of whether the reaction is nucleon-nucleon, $\gamma p$ or $\gamma \gamma$, being only a function of the cms energy $\sqrt s$.  This equality of survival probabilities is also independent of any particular factorization scheme, again depending only on the opacity of the eikonal being reaction-independent. The fact that our estimates of large rapidity gap survival probabilities are independent of reaction has interesting experimental consequences.

\section*{Acknowledgments}

This work was supported in part by the Department of Energy under contract
no.~DA-AC02-76-Er02289 Task D and by grants RFFI-98-02-17463 and NATO OUTR.LG 971390.

\end{document}